\documentclass[12pt]{article}

\usepackage{lineno}
\modulolinenumbers[5]
\usepackage{caption}
\usepackage[a4paper]{geometry}
\usepackage{amsfonts}
\usepackage{arydshln}
\usepackage{xtab}
\usepackage{setspace}
\usepackage{fancyhdr}
\usepackage[demo]{graphicx}
\usepackage[T1]{fontenc}
\usepackage[latin1,utf8]{inputenc}
\usepackage[main=english,italian]{babel}

\usepackage{sidecap}

\sidecaptionvpos{figure}{c}

\usepackage{type1ec}
\usepackage[final]{microtype}
\usepackage{amsmath,amssymb,amsthm,eucal,dsfont,bm,mathrsfs,stmaryrd}
\usepackage{lmodern}
\usepackage{textcomp}
\usepackage{pict2e,enumitem}
\usepackage{hyperref}
\usepackage[dvipsnames]{xcolor}
\usepackage[nodayofweek]{datetime} 
\usepackage{comment}

\hypersetup{
    pdfpagemode={UseOutlines},
    bookmarksopen,
    pdfstartview={FitH},
    colorlinks,
    linkcolor={black},
    citecolor={black},
    urlcolor={black}
            }

\definecolor{cornellred}{rgb}{0.7, 0.11, 0.11}

\usepackage{geometry}

\newdateformat{monthyear}{\monthname[\THEMONTH], \THEYEAR}

\renewcommand{\P}{\ensuremath{\mathbb{P}}} 

\theoremstyle{plain}

\theoremstyle{definition}

\renewcommand
        {\thefootnote}{\arabic{footnote}}
        
\newcommand{\symfootnote}[1]{%
\let\oldthefootnote=\thefootnote%
\stepcounter{mpfootnote}%
\addtocounter{footnote}{-1}%
\renewcommand{\thefootnote}{\fnsymbol{mpfootnote}}%
\footnote{#1}%
\let\thefootnote=\oldthefootnote%
}

\usepackage{tikz}
\usetikzlibrary{shapes.geometric}
\usepackage{floatrow}

\counterwithout{equation}{section}

\makeatletter
\def\bbibitem#1{\item[]%
    \if@filesw\immediate\write\@auxout{\string \bibcite {#1}{\the\value{\@listctr }}}\fi\ignorespaces}
\makeatletter

\title{What is so special about analogue simulations?}
\author{{Francesco Nappo}\footnote{Politecnico di Milano, Department of Mathematics, via Bonardi 9, Campus Leonardo, 20133, Milan (Italy). E-mail: \texttt{francesco.nappo@polimi.it}} \, \& {Nicolò Cangiotti}\footnote{Politecnico di Milano, Department of Mathematics, via Bonardi 9, Campus Leonardo, 20133, Milan (Italy). E-mail: \texttt{nicolo.cangiotti@polimi.it}}}
\date{}

\begin{document}

\maketitle

\begin{abstract}
This paper defends an account of terrestrial analogue simulations in black hole physics as instances of inferences from material analogy in science (Hesse 1963). We outline the main verdicts and recommendations deriving from this analysis, arguing that they not only fit the existing practice but are also more credible than those supported by prominent epistemological alternatives (e.g., Crowther et al. 2019, Dardashti et al. 2019).
\end{abstract}

\section{Introduction}
\label{Sec1}
Do black holes emit a form of photonic radiation analogous to heat? In a seminal paper, Hawking (1975) showed that semi-classical assumptions, i.e., approximations of quantum behavior in a relativistic framework, entail the prediction of black hole radiation. Specifically, modeling black holes by the Schwarzschild solution to Einstein’s equations, whose metric is:
\begin{equation}
\label{eq1}
ds^2=-\left( 1-\frac{2M}{R}\right)dt^2+\left(  1-\frac{2M}{R}\right)^{-1}dr^2+r^2\left(d\theta^2+\sin^2\left( \theta \right) d\phi^2\right),
\end{equation}
and taking a Hermitian scalar field operator $\phi$ obeying the covariant wave equation $\phi_{ab}g^{ab}=0$ in curved spacetime with metric $g$, a thermal emission near the event horizon can be predicted to occur. Strikingly, the resulting black hole temperature $T_H$ would be proportional to surface gravity . This observation allows for a redescription of a black hole in terms of  and horizon area A analogous to the relation between temperature and entropy in thermodynamics (Bekenstein 1972).
\smallskip

As the means for measuring black hole radiation currently escape us, the prediction of thermal radiation remains a theoretical possibility. While the resulting analogy of black holes with thermodynamic systems is suggestive, and indeed is often used as a benchmark for candidate theories of gravity, a ‘trans-Planckian problem’ remains. If, using Hawking’s model, we trace the stories of emitted photons back to their distant past, they appear to come out of energy regimes of wavelengths well below the Planck scale. Relativistic assumptions are expected to break down at those ultra-high energy regimes, in a way that puts Hawking’s own prediction into jeopardy.  
\smallskip

To overcome the impasse, Unruh (1981) considered a system in which quantized linear sound fluctuations (or ‘phonons’) play the role of massless scalar fields (photons) and the acceleration of a homogeneous fluid the role of a fixed but curved spacetime. The crucial observation is that phonons in the surrogate system obey equations mathematically the same as semiclassical equations for massless scalar fields in a fixed effective background geometry, viz: 
\begin{equation}
\label{eq2}
ds^2=-c^2dt^2+\left( dr \pm c\cdot \frac{r_0^2}{r^2}\,dt \right)+r^2\left(d\theta^2+\sin^2\left( \theta \right) d\phi^2\right),
\end{equation}
which leads, after a standard coordinate change, to the equation:
\begin{equation}
\label{eq3}
ds^2=-c^2\left( 1-\frac{r_0^4}{r^4} \right)d\tau^2+\left( 1-\frac{r_0^4}{r^4} \right)^{-1}dr^2+r^2\left(d\theta^2+\sin^2\left( \theta \right) d\phi^2\right).
\end{equation}
Part of the explanation is that the physics of phonons at ordinary energy regimes is equivalent to that of massless scalar fields in curved spacetime. Moreover, it can be shown that an accelerated fluid imposes a metric on the acoustic system whose features are completely determined by the velocity of sound and that of the intervening fluid (Unruh 1981). When the latter two quantities match, we have an acoustic horizon. Under an appropriate choice of radial velocity for the fluid, \eqref{eq3} is analogous to the Schwarzschild metric \eqref{eq1} used in Hawking’s model. Accordingly, one can re-write Hawking’s derivation to predict that sonic radiation occurs near a dumb hole’s horizon, with a temperature proportional to the rate of change of fluid velocity at the horizon:
\begin{equation}
\label{eq4}
T=\frac{1}{2\pi}\cdot \frac{dv}{dr}\Big |_{v=c}.
\end{equation}
Unruh’s insight that the fluid system – the ‘dumb hole’– could be obtained in terrestrial settings inspires an active research program in experimental physics (cf. Barcelo et al. 2019). Notably, tests on surface water waves indicate that sonic radiation may occur near a dumb hole’s acoustic horizon (Rousseaux et al. 2008, Weinfurtner et al. 2011).\footnote{This conclusion is reached indirectly, by simulating the sonic analogue of a white hole and inferring the radiation effect from the behavior of in-going modes and time-symmetry of physical laws (cf. Euvé et al. 2016).} Another important result is the report of spontaneous (i.e., not artificially stimulated) sonic radiation in Bose-Einstein condensates (Steinhauer 2016). The number of proposed realizations keeps growing (cf. Mertens et al. 2022).
\smallskip

While the fact that current analogue simulations are revealing new properties of the systems being tested and generating new physical ideas is beyond dispute, it remains controversial whether they can tell us anything about actual black holes. Unruh’s (1981) case for a positive answer rested on the observation that, just as semi-classical equations break down at ultra-high energies, so also by approximating \eqref{eq3} to the hydrodynamic limit we obtain the same ultra-high energies for emitted phonons. In the latter case, Unruh added, the atomic theory of matter prevents energy regimes with frequencies much lower than interatomic spacing. The dispersion relation, i.e., the relation between frequency and wavenumber, is accordingly different in an atomic fluid than in a continuum. Observing sonic radiation from a dumb hole could therefore indicate, by analogy, that the trans-Planckian problem does not represent as large of an obstacle for radiation as initially thought. 
\medskip

In this paper, we draw upon Hesse’s theory of material analogical inference and further developments (Hesse 1963) to articulate an epistemological account of simulations in black hole physics. One of our main conclusions will be that current analogue simulations only provide limited evidence for Hawking radiation. However, we will resist arguments according to which terrestrial simulations tell us nothing about black holes (Crowther et al. 2019) by showing that there are successful paths to the incremental confirmation of Hawking’s prediction which exploit locally robust physical connections.\footnote{Rovelli (2019) questions the relevance of philosophical notions of evidence and incremental confirmation to contemporary physics. We assume that the above notions are relevant to a reconstruction of the methodology of physics, yielding fine-grained methodological insights not capturable in more austere languages (see Sect. \ref{sec4}).} Furthermore, we will argue that the methodological recommendations for future confirmatory simulations following from our account are squarely more plausible than those which follow from Dardashti et al.’s (2019) prominent alternative. 
The discussion below is organized as follows. Section two briefly reviews the existing literature discussing the epistemology of analogue black hole simulations. Section three outlines and defends our novel approach. Section four highlights its main advantages. Section five concludes.

\section{Existing alternatives}
\label{sec2}
The current literature contains highly conflicting assessments regarding the epistemological status of black hole simulations. Notably, Dardashti et al. (2017) propose an account on which analogue simulations are a novel hybrid form of scientific inference -- taking off from a syntactic isomorphism between the models of distinct physical domains, yet capable of providing evidence for Hawking radiation. In a related paper, Dardashti et al. (2019) provide a Bayesian analysis of their notion of “confirmation via analogue simulation” (3). Conversely, Crowther et al. (2019) defend a skeptical position, claiming that terrestrial simulations do not provide any evidence and that seemingly confirmatory routes to Hawking’s prediction “beg the question” (2019:3705). 
\smallskip

Upon close inspection, some limits of both proposals emerge. In defense of their account, Dardashti et al. point to the availability of ‘universality theorems’ for Hawking radiation, namely analytic results establishing that a radiation effect follows invariably (i.e., no matter the microphysical details of underlying medium) from assuming a set of physical conditions of a general (and hence multiply realizable) kind.\footnote{Dardashti et al. (2019) additionally stress that black hole simulations involve “learning about the world by manipulating it” (2019:2) and “new evidence” (3). We take it, however, that many ordinary analogical inferences similarly involve manipulation and new evidence: e.g., learning about human beings from manipulating mice.} In Unruh and Schützhold’s (2005) version of the theorem, these conditions are: (a) there exists a freely falling frame for observers near the horizon; (b) the particles flowing out of the horizon are born in their ‘ground state’ with respect to freely falling observers; (c) the evolution of those particles is adiabatic. On Dardashti et al.’s (2017) interpretation, the observation of sonic radiation in terrestrial simulations, when paired with an appropriate universality result, provides evidence that “Hawking radiation \dots is independent of the details of the underlying micro-physics” (2019:4), and hence is robust to trans-Planckian issues.\footnote{Other results mentioned by Dardashti et al. (2017) include Barcelo et al.’s (2009) and Coutant et al.’s (2012).}
\smallskip

However, as pointed out by several authors (Crowther et al. 2019, Field 2022), the robustness claim that universality theorems establish is weak and conditional.\footnote{This is in tension with some of Dardashti et al.’s claims, e.g., that “the strength or quality of the inferences [via analogue simulations] is much greater than that of those via analogical reasoning” (68); cf. also Thebault (2019): “It is thus plausible to think of analogue experiments as prospective means for providing confirmatory support that is substantial [sic], rather than merely incremental” (312). These assessments are not justified by the consideration of the actual evidential affordance of universality theorems for Hawking radiation. In section four, we will show that inferences from simulations can confirm despite ultimately relying on ‘ordinary’ analogy.} Unruh and Schützhold’s result, for instance, would be evidence for the robustness of Hawking radiation if we had evidence that (a)-(c) hold for black holes. But, as Unruh and Schützhold (2005) note, counterexamples to (a)-(c) “do not appear to be unphysical or artificial” (11; cf. Gryb et al. 2020). The observation of sonic radiation in a terrestrial simulation can hardly be significant evidence that black holes satisfy (a)-(c). Consequently, it becomes questionable that “we can plausibly take the work of Unruh and Sch\"utzhold to be an argument for both the robustness and the universality of the Hawking effect” (Dardashti et al 2019:4). While universality holds, the argument for the robustness claim is weak.  
\smallskip

Crowther et al.’s argument that there cannot be any confirmation of Hawking radiation from an analogue simulation raises just as many doubts.  On one hand, one may interpret their claim that confirmation via analogue simulation “crucially rests on the assumption that quantum field theory (QFT) is applicable to [the target]” (3704) as the argument that, since QFT is not well-confirmed, and since Hawking’s derivations require QFT, terrestrial simulations cannot confirm Hawking radiation. However, the argument proves too much (cf. Lange 2019; Evans and Thebault 2020). When in \emph{Sidereus Nuncius} Galilei argued that the Moon has mountains and valleys, based on the similarities between the “wavy lines” (1610:53) observed on its surface and the shadows that mountains cast on valleys, the hypothesis that light behaves on the Moon just as it does on Earth was not well-confirmed, and yet was required to derive the prediction of valleys and mountains on the Moon. Applying Crowther et al.’s reasoning, we must rule out (implausibly) any confirmation of Galilei’s hypothesis from the similarities of lunar wavy lines with shadows.
\smallskip

On the other hand, Crowther et al. may be read as claiming that the lack of experimental access to a target prevents confirmation of a prediction about that target: as they write: “scientists cannot access T [the target], and so they cannot confirm if T and S [the source] \emph{actually are} in the same [\,\dots] class” (3709) and “in those cases of material models and simulations where the target system is experimentally accessible, the fact that they can yield confirmation is reflective of the ability of the models and simulations to be tested” (3724). However, the criterion of confirmability suggested by these passages is highly questionable. Observed properties of stars (e.g., the Cepheids’s temperature) are routinely used to predict yet unobserved properties of more distant stars, despite our inability to experimentally access them (cf. Lange 2004). Similarly, hypotheses about the human past are ordinarily supported by analogy with cultures familiar to us, despite our inability to experimentally access the past (Wylie 1985; Currie 2021). Considering these examples raise serious doubts on the tenability of Crowther et al’s criterion.
\smallskip

In summary, existing alternatives face important limitation. The evidential claim established by Dardashti et al.’s reconstruction is conditional and weak, raising the question whether inferences from analogue simulations should be construed as based upon universality theorems. Meanwhile, Crowther et al. have yet to articulate a credible epistemological criterion for why analogue simulations are in principle incapable of confirming. In what follows, we will argue that a third epistemological position has been neglected in this literature. Compared to existing rivals, our proposal not only yields a more realistic assessment of black hole simulations, but is also a source of more credible advice for assessing current simulations and designing new ones.

\section{Black hole simulations as material analogues}
\label{sec3}

A brief parenthesis about analogical inference in science will prove useful. A traditional view in epistemology distinguishes two senses of analogy in science: \emph{formal} and \emph{material} (Hesse 1963). The former refer to cases where there is a one-to-one correspondence between the properties and relations of two domains, but where the latter are otherwise unrelated: for instance, when one finds that notions of quantum mechanics can be applied to certain financial transactions (Hesse 1963:32). This sense contrasts with material analogy, in which source and target are related by more than isomorphism. For instance, extrapolations from mice to humans are often underwritten by an expectation of similarities in biological features, so that defeasible inferences can often rationally be drawn from, e.g., a given drug’s effects on mice to its similar effects on humans.
\smallskip

While Dardashti et al. view inferences from black hole simulations as based on formal analogy together with universality results, we defend the view that they can be material, instead, and ought to be evaluated accordingly. In seminal work, Hesse (1963) has posited the following conditions for a \emph{material analogy} between a source and a target system: 
\begin{enumerate}
\item {\em Materiality}: The similarities between the domains must be “material” or “pre-theoretic” (Hesse 1963:32), i.e., they must be cases of sharing of features that can be regarded as genuine respects of similarity \emph{before} and \emph{independently} of any analogical argument.
\item {\em Relevance}: the known similarities must be relevant to the occurrence of the predicted similarities (Hesse 1963:67), viz., it must be a serious possibility that the same causal or explanatory connection (stronger than correlation) between the properties of the source also holds between the known and the merely predicted properties of the target.\footnote{Hesse (1963:77) requires specifically \emph{causal connections} as relevance relations. Bartha (2009) argues that this is too restrictive: sometimes mathematical connections are at work. See also Nappo (2021) for discussion.}
\item \emph{No-Critical-Difference}: there must be no known or expected differences between source and target which directly and in the first instance undermine the possibility of analogical inferences from source to target (Hesse 1963:70).
\end{enumerate}
Recent work by Nappo (2022) shows that, when the above conditions are met, there can be incremental confirmation of a prediction about a target from observations in a material analogue.
\smallskip

In what follows, we assess the extent to which the conditions of material analogy apply to a \emph{prototypical} case of a terrestrial black hole simulation. This would involve the testing of an outgoing thermal spectrum effect in the vicinity of an artificial event horizon, where the test medium can be safely assumed to be homogeneous and with a dispersion relation that diverges from that described by classical dynamics at ultra-high energies.\footnote{The experiments that we target below are specifically those of the ‘classical’, stimulated variety, such as those on surface water waves. The reason is solely dialectical: if we can show that classical simulations can function as evidential tools, the verdict would be \emph{a fortiori} plausible for alleged simulations of the quantum variety.} Since the internal validity of current analogue experiments is subject to an on-going debate (cf. Steinhauer 2017), we take no stance below as to whether anything resembling the prototypical simulation has yet verified occurrences of sonic radiation. Our aim will be solely to assess whether black hole simulations may satisfy some independently plausible general conditions for fulfilling an evidential role.
\smallskip

In brief, our argument for the applicability of material analogy goes as follows. Terrestrial simulations resemble black hole horizons in broadly kinematic aspects: horizon stability and the mode of propagation of scalar waves in general relativity are similar to, respectively, artificially generated horizons and the way sound propagates in a fluid (Leonhardt \& Philbin 2008; Barcelo et al. 2009). At the same time, the specific processes whereby outgoing modes are created and emitted from horizons are physically dissimilar. However, this is not yet a defeater for the analogy, since the relevant mechanical aspects should not be read off from general relativity, but directly from a quantum theory of gravity. In this regard, critical differences between the behavior of horizons at extremes in the simulation and in black holes are not known; indeed, various results point to the possibility of locally robust connections between the real-world system and the analogue. Accordingly, terrestrial simulations can qualify as material analogues of black holes. 
\smallskip

In what follows, the key aspects of this reconstruction will be discussed in greater detail.
\subsection{Materiality}
\label{sec31}
First, a note about the condition’s scope. Dardashti et al. write that “analogue simulation \dots does not involve a material analogy in the sense of Hesse since there is not a \emph{physical property} common between target and source” (2019:2). They assume that material analogy requires that the same properties of the source are instantiated (to some extent) in the target. However, this reading is highly conservative. Hesse’s book contains a discussion of “analogue machines” (1963:92) in science, where the source is not readily available but “built in order to simulate the [target]” (92). As Hesse writes, these machines are “useful and necessary as predictive models precisely in those cases where the material substance of parts of the analogue is not essential to the model, but where the mutual relations of the parts are essential” (92). She adds that inferences from analogue machines may still count as material (“do not \dots provide counterexamples to the conditions suggested”, 92) despite the absence of physical properties common to both source and target. 
\smallskip

On a more liberal reading, \emph{Materiality} only requires that there be similarities among properties and relations of source and target that can be recognized \emph{before} and \emph{independently} of the analogical argument, in virtue of the language and assumptions that a scientific community shares (Hesse 1963:15).\footnote{Hesse (1963) herself considers the sound-to-light analogy as a classic instance of material analogy, claiming that their “similarities were recognized before any theory\dots for both sound and light was known” (1963:70).} To see how this condition applies, we must recall that the description of a black hole as a region of space where gravity is so strong as to entrap light does not depend on any specific physical theory – not even general relativity. When a relativistic framework is adopted, the analogy with sound being entrapped in a converging fluid is rather immediate. Unruh (2008) recounts using the analogy of a fish screaming inside a waterfall in a seminar in 1972 as an intuitive illustration of photons falling into a black hole. As Visser et al. (2002) confirm: “ideas along these lines have, to some extent, been quietly in circulation since the inception of general relativity” (2). Thus, at least some surface-level similarities between the systems satisfy the \emph{Materiality} requirement.
\smallskip

An easy rejoinder here is that the surface-level similarities between black holes and their terrestrial analogues are too faint to ground any reasonable inference to Hawking radiation. However, further material similarities can be noted. The most important ones are in broadly kinematic respects (Leonhardt \& Philbin 2008; Barcelo et al. 2009). On the one hand, the equivalence between the propagation of sound and light (in terms of scalar waves) was already noticed by Moncrief (1980), which led to the assumption of fixing velocity in developing the theory for both quantities. On the other hand, a long-lived horizon in the analogue system, by which the emission of radiation can be induced, is a kinematic phenomenon. Its stability allows for the fundamental characteristics of the horizon to emerge (in the same way as light-crossing time does for actual black holes). Although recognizing such additional similarities between black holes and their analogues requires physical theory, they are plausibly classified as ‘material’ when relativized to the accepted language and knowledge of the current physics community. 
\smallskip

The recognition of material similarities should not deter us from also noting material dissimilarities between the systems. There is a clear difference between the process whereby the spacetime geometry is generated: artificial black holes are created by means of moving media, but in general relativity the same role is played by the mass as stated by the Einstein field equations. Hence, the analogy can be affected by dynamical aspects that are not theoretically well described. This consideration partly motivates our assessment that current tests analogue simulations can only be limited evidence for predictions about black holes. From this perspective, there is no difference with other cases in which a prediction about a target system is supported by observations about a material analogue, where the two systems typically display both similarities and differences.
\smallskip

Importantly, in recognizing the above material similarities and dissimilarities between black holes and their terrestrial analogues we do not \emph{assume} that Hawking’s semi-classical model is “empirically adequate” (Crowther et al. 2019:3704) for black holes. Our assessment only requires that Hawking’s model is an adequate representation of black holes up to some degree of approximation (just as classical equations of fluid dynamics can be for ordinary fluids). To deny that this is a plausible assumption arguably diminishes the extent of our knowledge. The model’s classical aspects are based on general relativity, which is well-confirmed. The extra assumptions of QFT are relatively well-confirmed for flat spacetime; for curved spacetime, although the tests are less precise and conclusive, the theory remains a serious competitor. Section four will show, by a probabilistic argument, that these weak assumptions suffice for incremental confirmation.
\subsection{Relevance}
\label{sec32}

An important feature of Hawking's (1975) derivation is that it does not specify a physical process by which black hole radiation is emitted. Starting from the scalar field operator $\phi$, Hawking derives the expression for in-going particles using($a_i$) and creation ($a^{\dagger}_i$) operators: 
\[
\phi=\sum_i\left(f_ia_i+\bar{f}_ia^{\dagger}_i\right),
\]
where $f_i$ is a family of complex solutions of the wave equation $f_{ab}g^{ab}=0$. In a similar way, one can derive the expression for out-going waves and waves crossing the event horizon as:
\[
\phi=\sum_i\left(p_ib_i+\bar{p}_ib^{\dagger}_i+q_ic_i+\bar{q}_ic^{\dagger}_i\right),
\]
where $\{p_i\}$ and $\{q_i\}$ are sets of complex solutions standing for two respective wave equations. By equating the two expressions for $\phi$, one derives that annihilation operators for outgoing modes can be rewritten in terms of the operators for in-going modes as follows:
\[
b_i=\sum_i\left(\bar{\alpha}_{ij}a_i+\bar{\beta}_{ij}a^{\dagger}_i\right),
\]
which leads to writing the expectation value of the operator $b^{\dagger}_ib_i$ for the outgoing state as: 
\[
\langle0_{-} |b^{\dagger}_ib_i|0_{-} \rangle=\sum_i\left | \beta_{ij}\right |^2.
\]
Using the Fourier transform, Hawking (1975) proposed to side-step the complexity of computing the coefficients $\beta_{ij}$ via deriving an asymptotic form for them, which allows for arbitrarily high frequencies. He could then immediately estimate the number of particles emitted from the horizon, which is given by $(\exp(\pi \omega/\kappa)-1)^{-1}$ times the number of particles absorbed by the black hole. 
\smallskip

On the one hand, the absence of a specified mechanism for Hawking radiation raises several interpretational hurdles for theoretical physicists, particularly regarding its localization within the black hole or at the event horizon. On the other hand, the approximative nature of Hawking’s derivation emerges as a crucial aspect in making terrestrial simulations relevant. In Unruh’s model for an analogue black hole, one assumes that the fluid is sufficiently well approximated by the equations of continuum fluid dynamics, despite our knowledge that at the atomic scale – when high frequencies are in place – those equations lose meaning. The main reason why the similarities in broadly kinematic respects satisfy \emph{Relevance}, then, is that in both the black hole and in the simulation the respective physical models describing the systems entail a thermal effect via assumptions that we antecedently know to be unphysical, such as ultra-high energies. The parallels between the problem situations is arguably the main basis for increasing trust in Hawking radiation when sonic radiation is observed to be robust to ultra-high energies in terrestrial simulations.\footnote{As Unruh (1995) shows for the hypersonic flow case, numerical computations already yield that thermal emission from an analogue horizon does not depend on the ultra-high frequency allowed by the wave equations. }
\smallskip

A further result can be mentioned in this regard, which is not discussed by Crowther et al. and Dardashti et al. Unruh (2011) has shown that in a linear amplifier (such as a device that transforms an input signal into an output with different amplitude) quantum noise is thermal noise. Consider a simple model for amplifier, in which input and output are single modes denoted by $x$,$p$ and $Y$,$P$:
\[
Y=Ax \qquad \text{and} \qquad    P=Ap,
\]
and where $A$ is the amplification function (which is time-dependent). To allow for non-trivial amplification, we add a second input channel (an additional degree of freedom). Hence, we fix:
\[
Y=Ax+Bq \qquad   \text{and} \qquad     P=Ap+Er,
\]
where $r$ is the conjugate momentum to $q$. By demanding the standard commutation relation $[Y,P]=i\hbar$ for such a system, one obtains the following relations:
\[
E=-B   \qquad \text{and} \qquad    A^2-B^2=1,
\]
or equivalently:
\[
 A=\cos(\mu)    \qquad \text{and} \qquad     B=\sinh(\mu).
\]

An interesting result is that one can rewrite the system above in terms of creation and annihilation operators, inducing an instance of the so-called Bogoliubov transformations (Unruh 2011). In light of this, and after computations involving density matrices and the temperature of input channels, one can prove that the output thermal noise $T_{\text{Out}}$ due to vacuum fluctuations is:
\[
 T_{\text{Out}}=\frac{\omega_{\text{Out}}}{2\ln(\tanh(\mu))},
\]
where $\omega_{\text{Out}}$ is the output frequency. This result is significant because both amplification $\tanh(\mu)$ and frequency $\omega_{\text{Out}}$  are determined solely by the classical behavior of the amplifier. The argument generalizes to black holes insofar as, according to Hawking’s model, they are special cases of linear amplifiers. In more detail, starting from a semi-classical model, one can define a norm function (not necessarily positive) induced by the inner product between solutions $q_i'$, $\pi_i'$ and $q_i$, $\pi_i$:
\[
(q',q)=\frac{i}{2}\sum_i(\pi'_iq_i-q'_i\pi_i).
\]
If one then picks the solutions that have positive norm (thereby also determining the associated set of complex conjugate solutions) and combines them with standard annihilation and creation operators, one obtains quantum operators $Q_i$ and $\Pi_i$, which obey the standard commutation relation $[Q_i,\Pi_ j]=i\delta_{ij}$. In effect, this means that, given the classical solutions to the Hamiltonian, every conceivable quantum solution can be expressed as a linear combination of the classical ones.
\smallskip

In summary, our assessment is that the similarities in broadly kinematic respects satisfy the \emph{Relevance} requirement. Indeed, it is a serious possibility that the derivations used in Hawking’s and Unruh’s models are infinitary approximations of some actual physical connection between the physics of horizons and thermal effects – a connection that may be common to both source and target. We further highlighted the result that, even though what we test in a terrestrial simulation may be solutions to classical equations for the analogue horizon, the quantum radiation of the analogue system is also tested, in that it can be described as a linear combination of the classical solutions. This result does not give us full assurance that black holes will turn out to behave uniformly with terrestrial simulations. For further obstacles, of a kind we are not yet able to anticipate, may still interfere. However, the result that quantum noise for linear amplifiers is thermal noise reduces the gap between observing a classical thermal effect of an analogue horizon and reasoning to its quantum counterpart, thus supporting the relevance of the simulations.

\subsection{No-Critical-Difference}
\label{sec33}

The argument so far has been that, since it can be shown that the truncation of the continuum fluid equations at the atomic scale does not affect sonic radiation in the simulation, so we can expect that the truncation of the wave equations in the gravitational case (at whatever energy cutoff it may occur) does not affect the analogous prediction in black holes. However, a question still appears to be haunting our assessment. Tests on analogues may plausibly confirm that a thermal effect is robust to the breaking down of the continuum fluid dynamics equations at the atomic scale. But Hawking radiation is a quantum effect that must survive complications at a level of description of gravity that we entirely ignore.\footnote{ Among other things, Hawking’s (1975) derivations assume that there is no back-reaction between the effective background geometry and the quantum field -- something that we have reason to regard as a mere idealization.} In particular, we ignore the mechanism whereby waves are regularized away from ultra-high energy regimes. This observation may well seem to preclude epistemically responsible inferences about black holes from terrestrial analogues. In Hesse’s language, it seems that an “essential difference” (1963:78) emerges upon close inspection. 
\smallskip

The possible presence of critical differences is admittedly the hardest issue for the evidential status of black hole simulations. While attempts could be made to escape the difficulty via purportedly testing quantum versions of thermal radiation (Steinhauer 2016), here we will point to the availability of a response of a weaker but more general kind – one that does not rest on contested interpretations of experimental settings (cf. Steinhauer 2017).\footnote{Cf. Hangleiter et al. (2022, ch. 5) for a discussion of quantum analogue simulations. Their take on the epistemic import of such simulations (in ch. 7) is in line with Dardashti et al.’s reconstruction (discussed in section two).} It is enclosed in Unruh’s (2014) claim that, even though “something could make the gravitational system behave differently from any analogue system [\,\dots] it is hard to imagine what that something could be” (544). 
\smallskip

To elaborate, the mechanism of regularization of extreme waves at the horizon in analogue systems is the strongly dispersive regime that characterizes those systems. Conversely, the physics describing the dynamics of the waves when approaching the black hole horizon, and specifically the mechanism for regularizing the extreme of waves at horizons, is unknown. Although the spacetime geometry due to the gravitational field is well defined (in terms of Schwarzschild coordinates), waves are trapped oscillating with decreasing wavelengths, and they develop a logarithmic phase singularity that induces an exponentially long escape time for the rays close to the horizon. Therefore, there is an open question about whether, and if so how, an analogue of the dispersion mechanism regularizes the extremes of waves at the horizon for actual black holes. 
\smallskip

Still, some remarkable aspects of continuity must be noted. First, it is known that, as we move away from strongly dispersive regimes towards moderate ones, analogues of Hawking radiation still occur. For very low dispersion regimes, numerical simulations and physical tests produce more ambiguous results, for reasons that are not well-understood (Leonhardt and Philbin 2008). Secondly, it has been shown analytically that there are speculative mechanisms of regularization of extreme waves based on known physical patterns and alternative to dispersion which still demonstrate Hawking radiation (e.g., Brout et al. 1995). Hence, the assumption that the actual mechanism of wave regularization is not dispersion does not prevent Hawking radiation from occurring.  Overall, Hawking radiation appears to be remarkably robust to a variety of underlying dynamical assumptions, which gives further confidence in its applicability to black holes.
\smallskip

In summary, our assessment is that, despite our ignorance of the underlying dynamics, no critical differences are yet known or expected between black holes and their terrestrial analogues. Admittedly, the waves of interest for Hawking effects oscillate between two different physics (so to speak), one coming from the kinematics description providing the geometric framework, namely general relativity, and the other that we expect from the dynamics of in the microscopic high-energy description, namely quantum mechanics. A unified theory of gravity and quantum mechanics is a central open question in the foundations of physics. However, the study of regularization of waves extreme at the horizon do not yet give signs of deep discontinuities as we approach lower dispersion regimes or alternative regularization mechanisms for waves. 
Ultimately, our assessment is that inferences from black hole simulations are analyzable as arguments from material analogy. The next section highlights some advantages of this analysis.

\section{Discussion}
\label{sec4}

To summarize, the previous section has offered an assessment of inferences from black hole simulations as candidate instances of Hesse’s category of material analogical inferences in science. Compared to other instances (e.g., extrapolations from mice to humans in medicine), black hole simulations are distinguished by a relatively small set of known material similarities and absence of a known common causal mechanism for the generation of the relevant radiation effect. For these reasons, we find that the evidence for Hawking radiation that current analogue simulations can provide is very limited. Despite this, our assessment suggests that the conditions for material analogy, and particularly the No-Critical-Difference condition, are met. A notable consequence of this view is that, in addition to offering new insights about the systems under test, there can be successful paths to the confirmation of Hawking’s prediction from a terrestrial simulation.
\smallskip

The stylized Bayesian representation of confirmation by material analogy developed in Nappo (2022) corroborates the informal assessment. Consistent with that framework, we consider the relation between the following three statements as entertained by an ideal Bayesian agent:
\begin{itemize}
    \item[(T)]	A radiation effect occurs in actual black holes.
    \item[(E)] An occurrence of sonic radiation in a terrestrial simulation is observed.
    \item[(B)] Within a range of variation that includes black and dumb holes, the radiation effect is robust to details at the level of the system’s actual physical realization.
\end{itemize}
The latter statement, B, plays the role of ‘bridge hypothesis’, potentially connecting evidence gathered in a source to predictions made about a materially analogous target (cf. Nappo 2022).
\smallskip

The following probabilistic conditions can be shown to suffice for T’s confirmation by E in a Bayesian sense (i.e., for $\P(T\, |\, E) > \P(E)$):
\begin{itemize}
    \item[(C1)] $ 0 < \P(T)$, $\P(E)<1$;
    \item[(C2)] $0< \P(B) < 1$, where: 	
    \begin{itemize}
        \item[(a)] $\P(T \,| \,B) > \P(T)$;
        \item[(b)] $\P(T \,|\, E\, \&\, B) \ge \P(T \,|\, B)$;
        \item[(c)] $\P(T \,| \,E \, \&\, \neg B) \ge \P(T \,|\, \neg B)$;
    \end{itemize}
    \item[(C3)] $\P(E \, |\, B) > \P(E \, |\, \neg B)$.
\end{itemize}

C1-C3 are plausibly satisfied for a black hole simulation as described above. C1 is unproblematic. C2(a) is plausible if we consider that it is at least a serious epistemic possibility that a semi-classical model may be a suitable approximation for the physics of black holes in the respects that are relevant to $T$’s obtaining. C2(b) and (c) are weak transitivity assumptions, requiring merely that the observation of sonic radiation does not disconfirm $T$ if $B$ is known or (alternatively) if $\neg B$. In particular, C2(c) can be accepted on grounds that E is irrelevant to $T$ if $\neg B$ is known (i.e., if we are certain that the micro-physics at ultra-high energies affects the radiation effect); hence, $\P(T \,| \,E \,\&\, \neg B)=\P(T\,| \,\neg B)$. Finally, since there are similarities between black holes and terrestrial simulations which are both material and relevant, and no critical difference is yet known, it is plausible that $\P(B \,| \,E)$ is ever slightly greater than $\P(B)$, which is equivalent to C3. 
\smallskip

The analysis of black hole simulations as material analogues provides a sober assessment of their epistemological status in current black hole physics. As discussed, the limited set of known material similarities and absence of a known common mechanism poses an important limit to the evidential role of analogue simulations. The latter can accordingly integrate, but not replace, more direct forms of evidence for Hawking radiation. This verdict is broadly in line with that of Unruh (2008) and other proponents of analogue experiments, who recognize that the method is currently limited by the large uncertainties regarding black holes and their underlying physics.\footnote{Various aspects of our proposal are compatible with discussions in Sterrett (2017), Leonhardt \& Philbin (2008).} It is also in line with one interpretation of Dardashti et al.’s (2019) verdict, according to which analogue simulations can confirm incrementally (albeit not ‘substantially’ as Thébault 2019 argues) Hawking’s prediction. Our Bayesian analysis has the similar implication that $\P(T\, | \,E)> \P(T)$.
\smallskip

While the verdict is similar, the epistemological story behind the two accounts are sharply different. Dardashti et al.’s proposal relies upon the availability of universality results which, in combination with the observation of sonic radiation in a variety of analogues, supposedly yield evidence for Hawking radiation. Notably, the confirmatory path identified by our account does not rest on universality considerations (which, as discussed in section two, do not strengthen the inferences) but rather on the identification of locally robust connections. There may be, in other words, no way to generalize what physical conditions black and dumb holes have in common into a precisely definable ‘universality class’; a vaguer yet significant similarity in the relevant physical conditions, when paired with a judgment that there is no known critical difference, may suffice to confer evidential force to the inferences. In this sense, the confirmation is distinctively analogical and does not require sameness of the respective modelling frameworks under a given description.
\smallskip

Speaking of the reasons to favor our account, we think that our analysis better captures the nature of the reasoning involved. One issue with Dardashti et al.’s view is that, in the absence of a relation of material analogy between black holes and their terrestrial analogues, observations made in the simulations seem to constitute hardly more than enumerative evidence for the robustness of Hawking radiation. That is to say, the observation of radiation effects in the source is viewed as evidence for Hawking radiation only by way of belonging to the same postulated class of phenomena. However, enumerative induction is known to be especially weak when there is antecedent knowledge of differences between observed and predicted instances, as in this case. Accordingly, we believe that it is false that inferences from analogue simulations are stronger when construed along the lines that Dardashti et al. propose than as arguments based upon similarity between the relevant systems (cf. Dardashti et al. 2017:68). However, since justifying this claim further would lead us into some deep waters, we will not focus on it in what follows.
\smallskip

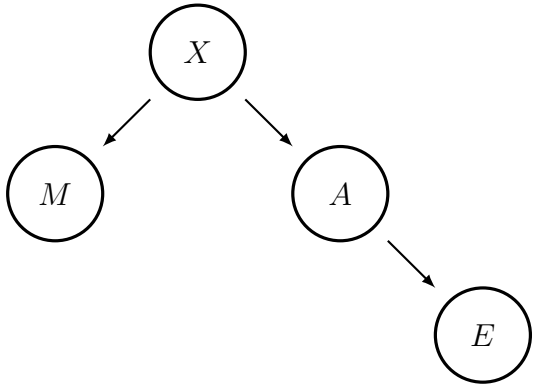
\begin{SCfigure}
\centering
\begin{tikzpicture}[scale=1.15,>=latex]
\draw[black, very thick] (5,5) circle[radius=0.5cm];
\draw[black, very thick] (3.5,3.5) circle[radius=0.5cm];
\draw[black, very thick] (6.5,3.5) circle[radius=0.5cm];
\draw[black, very thick] (8,2) circle[radius=0.5cm];
\node[scale=1] at (5,5) {$X$};
\node[scale=1] at (8,2) {$E$};
\node[scale=1] at (3.5,3.5) {$M$};
\node[scale=1] at (6.5,3.5) {$A$};
\draw[thick,->](4.5,4.5)--(4,4);
\draw[thick,->](5.5,4.5)--(6,4);
\draw[thick,->](7,3)--(7.5,2.5);
\end{tikzpicture}
\caption{\small{Dardashti et al.’s (2019) Bayesian network analysis for arguments from analogue simulations. The arrows between nodes in the model express relations of probabilistic dependence: for instance, the arrow from $X$ to $A$ stands for $\P(X\, |\, A)>\P(X)$. The absence of an arrow between two nodes expresses conditional independence: for instance, the absence of an arrow from A to M stands for $\P(M \,|\, A \, \& \, X)=\P(M \,|\,X)$.}}
\label{Fig1}
\end{SCfigure}

Another advantage over Dardashti et al.’s view that we shall rather stress here concerns the capacity of the proposed account to offer sound methodological insight. As a telling illustration, it is worth focusing on a specific claim by Dardashti et al.’s (2019), viz., that physicists ought to look for sonic radiation in media whose physics is not well-known rather than on media currently being used, whose physics is well-known (11). This recommendation of Dardashti et al.’s framework stems from their reconstruction of inferences from analogue simulations as based on universality results. On their Bayesian analysis, evidence of sonic radiation E confirms Hawking’s semi-classical model only by way of confirming $A$, that a model isomorphic to Hawking’s is empirically adequate for dumb holes (Fig. \ref{Fig1}). In turn, $A$ confirms the empirical adequacy of Hawking’s model ($M$) when universality theorems vouch for the robustness of the radiation effect to ultra-high energy physics ($X$). It follows from this analysis that $E$ confirms $M$ to a much greater extent when $\P(A)$ is low, i.e., when one ignores the physics of the test medium, than when $\P(A)$ is high. 
\smallskip

In our view, Dardashti et al.’s recommendation does not constitute sound advice. Consider, for illustration, a recent proposed realization of analogue Hawking radiation, by Mertens et al. (2022). The authors introduce a condensed matter analog in the form of a one-dimensional tight-binding model, which involves constructing an electronic band structure by approximating the Hamiltonian of the system. In this setting, it is possible to create an analog of a gravitational horizon by a quench of a homogeneous system into one with particular position-dependent hopping parameters (Morice et al. 2021). Evidently, no meaningful test of radiation could be achieved in this setting without 
prior knowledge of the properties of the system under investigation. Among other things, Mertens et al. argue that the number of particles measured by a static observer before and after formation of the horizon in a one-dimensional lattice model indicates a thermal distribution of particles. Without knowledge of the underlying physics, however, there would be no basis for such claim.

Insofar as it treats analogue simulations as continuous with the methodology of material analogical inference, the proposed analysis does not have the same implication. According to conditions C1-C3 above, $B$ (and, by transitivity, $T$) is more confirmed by E the larger $\P(E \,| \,B)$ is than $\P(E \,| \, \neg B)$. This can be interpreted by appealing to standard criteria for when the conclusion of an inference from material analogy is strengthened (cf. Hesse 1963; Wylie 1985). One is by increasing what one might call the ‘severity of the test’: we learn more the closer the conditions of the source are to those of the target system. This is obvious if one considers everyday examples: we are more confident that a given drug will be effective on humans after testing it on mice than on lizards. Similarly, if dumb holes come closer to replicating the alleged conditions of Hawking radiation, then $\P(E \,|\, B)- \P(E \,| \,\neg B)$ will be comparably larger, resulting in stronger confirmation.
\smallskip

Secondarily, one can strengthen the conclusion of a material analogical argument by expanding the ‘data points’ (Currie’s 2016:89 expression), i.e., by considering whether the relevant effect holds in a variety of diverse analogue realizations. Again, this is plausible advice in many ordinary examples: observations of a drug’s response in a variety of mice populations (as opposed to a specific variety) typically strengthen the case for the drug’s efficacy on humans. Similarly, from the model C1-C3 above we can easily derive that, if ‘$E$’ and ‘$F$’ term two analogue realizations of sonic radiation, $\P(E \, \&\, F \,|\, B)$ will plausibly be larger than $\P(E \,|\, B)$, and hence there will be more confirmation.\footnote{ It does not follow from our account that “we can perform a large number of terrestrial analogue experiments and thereby achieve an ever increasing\dots degree of confirmation.” (Crowther et al. 2019:3722) This would falsely assume that each observation of sonic radiation in the analogue system counts the \emph{same} towards confirmation.} Importantly, placing a confirmatory premium on the diversity of the realizations is not equivalent to placing a premium testing on unknown or exotic media, as in Dardashti et al.’s (2019) account, since diverse realizations can be obtained by tests on a variety of \emph{familiar} media.
\smallskip

The above example aims to provide a concrete illustration of how, by treating inferences from analogue simulations as analogical arguments in the material mode, the proposed account is in a privileged position to not only offer fine-grained verdicts regarding different analogue simulations, in recognizing their different potential for informativeness to cosmology, but also serve as a basis for credible methodological recommendations for scientific practice. To clarify, our claim is not that the above advice is more credible than Dardashti et al.’s because it is intuitively more plausible. Rather, the recommendations are more credible because they are grounded in a well-established methodology of material analogical inference in science. The advice is therefore more \emph{principled}, as it is consistent with a general account of what makes analogical arguments in science confirmatory; it is also more \emph{trustworthy} as it relies on a well-documented record of applying analogical reasoning in science, rather than on a methodology built specifically for a case-study. 

\section{Conclusion}
\label{sec5}
This paper has outlined an epistemology for black hole simulations in physics based upon Hesse’s (1963) general framework of material analogy in science. Our assessment is that such simulations may provide only limited evidence for Hawking radiation. However, we have resisted arguments according to which terrestrial analogue simulations tell us nothing whatsoever about black holes (Crowther et al. 2019) by showing that there are successful paths to the confirmation of Hawking’s prediction which exploit locally robust connections among physical systems. These paths are notably different from those which exploit isomorphisms paired with universality theorems, as in Dardashti et al. 2017. We have further argued that our account offers recommendations to physicists aiming to assess current experiments and design new ones which are squarely more plausible than those offered by Dardashti et al. (2019). Applications of the proposed framework to the methodological appraisal of specific proposed simulations will be the subject of future work.

\end{document}